%% file: main.tex
\documentclass[%
reprint,
 amsmath,amssymb,
 aps,
]{revtex4-2}

\usepackage{amsmath,amssymb, bm}
\usepackage{graphicx}
\usepackage{dcolumn}
\usepackage{bm}
\usepackage{physics}
\usepackage{comment}
\usepackage{ulem}
\usepackage{xcolor}
\usepackage{hyperref}

\usepackage{xifthen}
\newcommand{\vq}[2][]{                
  \ifthenelse{\isempty{#1}}           %
    { \hat{\pmb{#2}} }                
    { \hat{\pmb{#2}}_\mathrm{#1} }    
}
\newcommand{\vqtil}[2][]{             
  \ifthenelse{\isempty{#1}}           %
    { \hat{\til{\pmb{#2}}} }                
    { \hat{\til{\pmb{#2}}}_\mathrm{#1} }    
}
\newcommand{\tq}[2][]{                
  \ifthenelse{\isempty{#1}}           %
    { \mathbb{#2} }                   
    { \mathbb{#2}_\mathrm{#1} }       
}
\newcommand{\vs}[2][]{                
  \ifthenelse{\isempty{#1}}           %
    { \mathbf{#2} }                   
    { \mathbf{#2}_\mathrm{#1} }       
}
\newcommand{\matr}[4]{
  \begin{bmatrix}
    #1 & #2 \\
    #3 & #4
  \end{bmatrix}
}

\newcommand{\vect}[2]{
  \begin{bmatrix}
    #1  \\
    #2 
  \end{bmatrix}
}

\newcommand{\Tq}[3][]{                
  \ifthenelse{\isempty{#1}}           %
    { \mathbb{#3} }                   
    { \mathbb{#3}^{\rm #1}_{\rm #2} }       
}

\newcommand{\mr}[1]{\mathrm{#1}}

\date{\today}

\begin{document}

\preprint{APS/123-QED}
\title{Quantum Back Action Evasion with Reservoir Engineering}
\author{Yohei Nishino$^{1,2}$}
 \email{yohei.nishino@grad.nao.ac.jp}
\author{James W. Gardner$^{3,4}$}
\author{Yanbei Chen$^{4}$}
\author{Kentaro Somiya$^{5}$}

\affiliation{$^1$Department of Astronomy, University of Tokyo, Bunkyo, Tokyo 113-0033, Japan,}
\affiliation{$^2$Gravitational Wave Science Project, National Astronomical Observatory of Japan (NAOJ), Mitaka City, Tokyo
181-8588, Japan,}
\affiliation{$^3$OzGrav-ANU, Centre for Gravitational Astrophysics, Research Schools of Physics, and of Astronomy and Astrophysics, The Australian National University, Canberra, ACT 2601, Australia}
\affiliation{$^4$Walter Burke Institute for Theoretical Physics, California Institute of Technology, Pasadena, California 91125, USA} 
\affiliation{$^5$Department of Physics, Institute of Science Tokyo, Meguro, Tokyo, 152-8551, Japan} 
\date{\today}

\begin{abstract}
We propose a back-action evading scheme for a free mass that combines reservoir engineering with velocity measurement. The underlying principle follows the double-pass-type speed meter, which measures the mirror's velocity using a nonreciprocal interaction. In our method, the nonreciprocal coupling is realized through reservoir engineering, following the recipe proposed in [Phys. Rev. X 5, 021025]. We show that reservoir engineering can reproduce the double-pass speed meter with optimal feedforward, using only reciprocal interactions. The resulting force sensitivity surpasses the standard quantum limit, providing an alternative route to quantum back-action evasion in cavity optomechanical systems.
\end{abstract}
\maketitle
\input{chap_main}

\bibliography{ref} %

\end{document}

%% file: chap_main.tex
\section{Introduction}

Cavity optomechanics focuses on the interaction between light confined to a resonator (e.g.\ an optical or microwave cavity) and a mechanical element (such as a movable mirror, membrane, or nanoscale oscillator)~\cite{RevModPhys.86.1391}. The radiation pressure exerted by the intracavity electromagnetic field provides the optomechanical coupling mechanism and allows the exploration of mechanical systems in the quantum regime: Numerous studies have focused on cooling mechanical oscillators to their motional ground state~\cite{PhysRevLett.83.3174, PhysRevLett.98.030405,Chan_2011, Teufel_2011}, investigating the interface between classical and quantum physics~\cite{Chen_2013}, and controlling quantum states for quantum information processing~\cite{doi:10.1126/science.1244563,Safavi_Naeini_2011}. The precise measurement of extremely weak forces acting on a mechanical oscillator has attracted particular attention in the context of quantum metrology~\cite{RevModPhys.52.341, 1995qume.book.....B, RevModPhys.82.1155}.

When an external force is applied to the mechanical oscillator, its displacement modulates the electromagnetic field in the cavity. In the linear regime, the force or mirror position can thus be sensed with high precision from measuring the modulation. At the same time, radiation pressure can also enhance the oscillator’s response to external forces and improve the measurement precision, e.g.\ through an optical spring~\cite{RevModPhys.86.1391,Chen_2013}. However, because radiation-pressure fluctuations themselves introduce quantum noise, one encounters a trade-off between quantum shot noise (which decreases with increasing optical power) and back-action noise (which increases with optical power).

Braginsky demonstrated that this trade-off leads to the standard quantum limit (SQL) on the precision of weak force sensors~\cite{1968JETP.26.831}. Caves later examined ponderomotive squeezing and clarified the disturbance of mirror by the radiation pressure noise in interferometers~\cite{PhysRevLett.45.75}. Since then, researchers have proposed several methods to mitigate back-action noise, including squeezed light injection~\cite{Meystre1983QuantumOE, PhysRevD.23.1693, PhysRevD.30.2548, Jaekel_1990}, variational readout~\cite{PhysRevD.65.022002, VYATCHANIN1995269}, negative-mass systems~\cite{Julsgaard_2001, PhysRevLett.102.020501, PhysRevA.87.063846, PhysRevLett.105.123601, PhysRevLett.117.140401, M_ller_2017, PhysRevLett.121.031101}, and optical-spring techniques~\cite{PhysRevD.65.042001}. Among these, the concept of measuring a quantum non-demolition (QND) observable—one that commutes with itself at different times—has been studied extensively~\cite{doi:10.1126/science.209.4456.547}. For a free mass, its momentum (proportional to its velocity) is a QND observable and the corresponding device that measures the momentum is called a speed meter~\cite{PhysRevD.61.044002, BRAGINSKY1990251}.

In this paper, we propose a new method to realize back-action evasion via a speed meter. Our approach is inspired by two key ideas: (1) the concept of double-pass speed meters developed in Ref.~\cite{ BRAGINSKY1990251, 2018LSA.....7...11D} for gravitational-wave detection~\cite{PhysRevLett.22.1320, Saulson:262163,Sathyaprakash_2009} and (2) the reservoir-engineering framework for generating nonreciprocal interactions introduced by Metelmann and Clerk~\cite{PhysRevX.5.021025}. We show that reservoir engineering can realize the functional equivalent of a double-pass speed meter and thereby reduce back-action noise. As the scheme relies solely on reciprocal cavity interactions, it eliminates the need for polarization circulators or Sagnac-type interferometers, which would otherwise require modifications to the interferometer infrastructure or optical properties. This method offers an alternative route to surpass the SQL when measuring weak forces in cavity optomechanics.

The rest of this paper is structured as follows. In Section~\ref{sec:II}, we discuss the principle of back-action evasion by means of velocity measurement. Section~\ref{sec:III} presents the reservoir-engineering method for producing nonreciprocal interactions. Then, in Section~\ref{sec:IV}, we show the force sensitivity of the velocity measurement comparing with a position measurement. Finally, we discuss possible applications in Section~\ref{sec:discussion} and conclude in Section~\ref{sec:conclusion} with some remarks on future research directions.

\section{Back-action Evasion via Velocity Measurement}\label{sec:II}

In this section, we briefly outline the speed meter principle (see also, e.g., Chapter~IV of Ref.~\cite{1995qume.book.....B} or Section~4.5 in Ref.~\cite{Danilishin_2012}). 

The momentum of a free mass is a QND observable:
\[
[\hat{p}(t),\ \hat{p}(t+\tau)] = 0.
\]
In contrast to its position which is not:
\[
[\hat{x}(t),\ \hat{x}(t+\tau)] \neq 0.
\]
This means that while position measurements will generate back-action at later times, momentum measurements will not. In principle, this means that we can measure the momentum with arbitrarily high precision. Since the momentum of a free mass is proportional to its kinetic velocity, i.e.\ $\hat{p}(t)=m\hat{v}(t)$ for a mirror mass $m$, this momentum meter is also called a ``speed meter".

More precisely, when the mechanical object is coupled to a probe such as light, the canonical momentum \(\hat{p}\) is no longer directly proportional to the velocity \(\hat{v}\); instead, it takes the form
\[
\hat{p} = m\hat{v} - \alpha\hat{a}_1,
\]
where \(\hat{a}_1\) denotes the amplitude quadrature of the probe light and \(\alpha\) is the constant optomechanical coupling strength. This back-action term ($- \alpha\hat{a}_1$) can be canceled due to the ponderomotive correlation between the amplitude \(\hat{a}_1\) and phase \(\hat{a}_2\) quadratures of the outgoing field (see Section~2.11 of Ref.~\cite{HaixingPhD2010}). This thus allows us to surpass the SQL.

\begin{figure}
    \centering
    \includegraphics[width=0.9\linewidth]{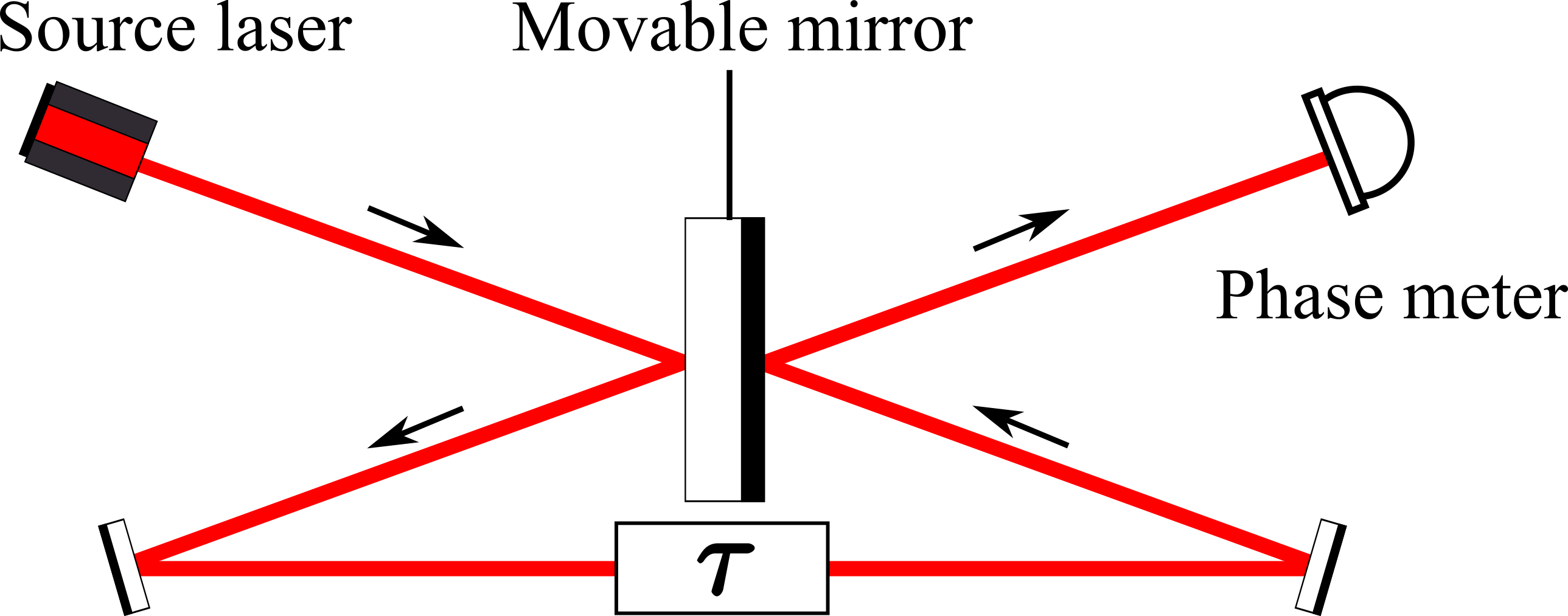}
    \caption{A simplified model of the nonreciprocal speed meter presented in Ref.~\cite{Danilishin_2012}.}
    \label{fig:Speedmeter}
\end{figure}

A speed meter must satisfy two key criteria: (1) the probe light must interact coherently with the test object on two distinct occasions and (2) the respective interaction Hamiltonians governing the probe-object interactions must have opposite signs~\cite{PhysRevD.61.044002, 2018LSA.....7...11D}. These features are illustrated in Fig.~\ref{fig:Speedmeter}. In the shown scheme, probe light from the source laser first strikes the left surface of the object mirror. The reflected light is then recycled to the back side of the mirror, where it impinges on the mirror for a second time. At each reflection, the mirror's displacement, e.g.\ at times $t$ and $t+\tau$, is imprinted on the phase of the light but with opposite sign. The total phase is thus given by
\begin{align}
    \phi \propto \hat{x}(t+\tau) - \hat{x}(t) \sim \tau\,\bar{v},
\end{align}
where \(\bar{v}\) denotes the mean velocity of the mirror. By measuring this phase using a phase meter (e.g., via homodyne detection), the measurement back-action is effectively canceled, as the forces imparted during the two interactions act in opposite directions.

The cascaded scheme illustrated in Fig.~\ref{fig:Speedmeter} involves a simple kind of nonreciprocal interaction: the light incident on the back of the mirror does not return to interact with the front of the mirror again. In the gravitational-wave detection community, this concept was first proposed for resonant-type bar detectors, and later several implementations for interferometric detectors have been proposed~\cite{2018LSA.....7...11D, nishino2025teleportationbasedspeedmeterprecision}. 
In this work, we focus on engineering this nonreciprocal interaction in a different way, as explained below.

\section{Nonreciprocal Interaction via Engineered Reservoir}\label{sec:III}

In this section, we realize the nonreciprocal speed-meter interaction through reservoir engineering~\cite{PhysRevX.5.021025} combined with conditional feed-forward.

\begin{figure}
    \centering
    \includegraphics[width=1.0\linewidth]{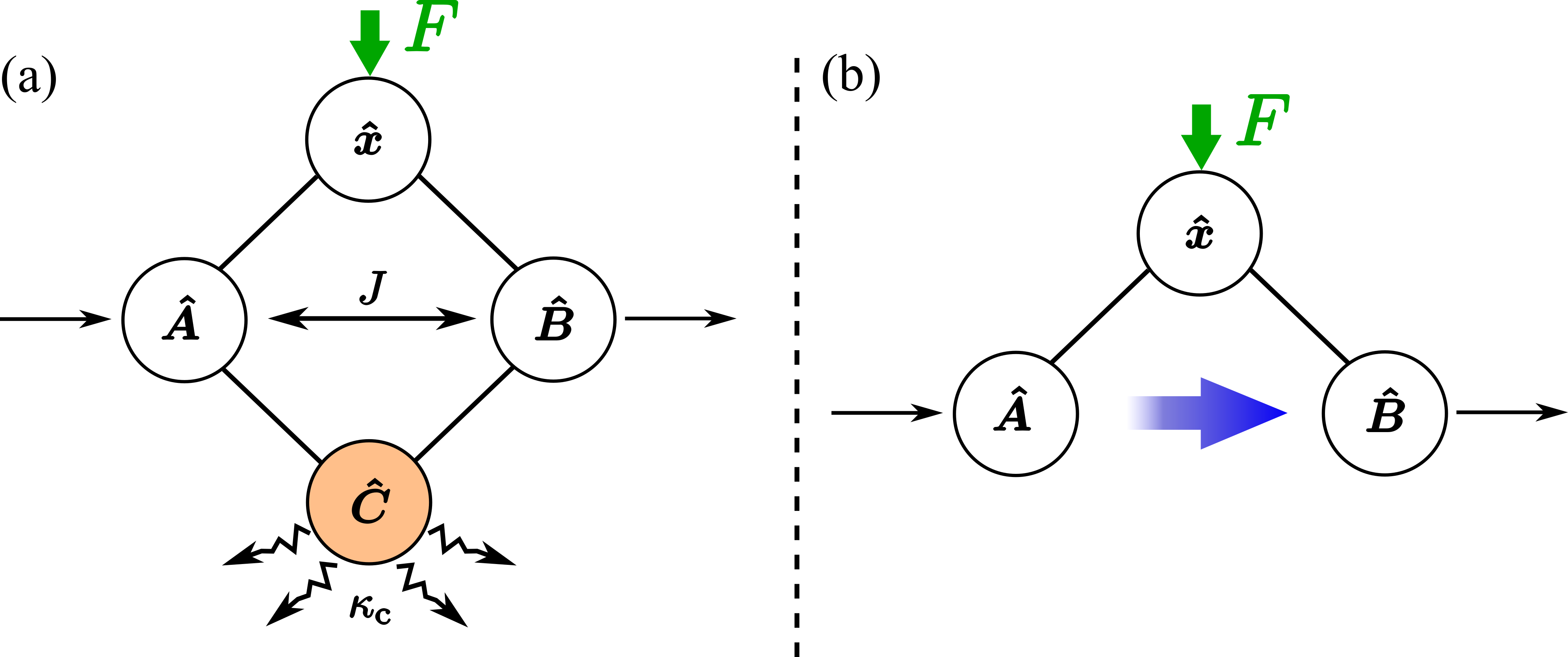}
    \caption{(a) Mode diagram of the speed meter with an engineered reservoir. Two cavity modes $\hat{A}$ and $\hat{B}$ are coupled to a mechanical mode $\hat{x}$ experiencing an external force. Modes $\hat{A}$ and $\hat{B}$ are coupled via a beam-splitter interaction with coupling rate $J$, while the auxiliary mode $\hat{C}$ is strongly damped and coupled separately to $\hat{A}$ and $\hat{B}$. (b) Mode diagram of the nonreciprocal speed meter. By tuning the parameters in (a), one can effectively create nonreciprocal coupling between cavities $\hat{A}$ and $\hat{B}$, equivalent to the double-pass--type speed meter described in Ref.~\cite{2018LSA.....7...11D}.}
    \label{fig:Engineered_reservoir}
\end{figure}

We consider the mode diagram shown in Fig.~\ref{fig:Engineered_reservoir}a, which consists of four modes. Here, $\hat{x}$ denotes the position operator of the mechanical mode that interacts with two cavity modes, $\hat{A}$ and $\hat{B}$. An auxiliary cavity mode $\hat{C}$ acts as an engineered reservoir and is strongly damped. The cavity modes $\hat{A}$ and $\hat{B}$ are coupled via a beam-splitter interaction, described by the Hamiltonian
\[
\hat{\mathcal{H}}_\mathrm{coh} = - \hbar J\, \hat{A}^\dagger \hat{B} + \mathrm{H.c.},
\]
where $J$ is the coupling coefficient. In addition, the interaction between the modes $\hat{A}$, $\hat{B}$, and the reservoir mode $\hat{C}$ is described by another beam-splitter Hamiltonian with a complex coefficient $J^\prime$:
\[
\hat{\mathcal{H}}_\mathrm{SB} = -\hbar J^\prime\, \hat{C}^\dagger (\hat{A} + \hat{B}) + \mathrm{H.c.}
\]
(SB stands for system-bath coupling).
For a more detailed derivation based on a master equation approach, see Ref.~\cite{PhysRevX.5.021025}. The interaction between the mechanical mode and the cavity modes is given by
\begin{align}
\hat{\mathcal{H}}_\mathrm{main} = -\hat{x}F + \hbar \alpha \hat{x}\bigl[(\hat{A}+\hat{A}^\dagger) - (\hat{B}+\hat{B}^\dagger)\bigr],
\end{align}
where $F$ is a weak external force and $\alpha$ is the optomechanical coupling strength. We explicitly assign a minus sign to $\hat{B}$ to satisfy the velocity-measurement condition; however, this phase can be tuned via other interaction Hamiltonian such as $\hat{\mathcal{H}}_\mathrm{coh}$ or $\hat{\mathcal{H}}_\mathrm{SB}$.

We work in a rotating frame where the three cavity modes are resonant, making the interaction Hamiltonian time-independent. In addition, the free evolution of the mechanical oscillator is described as:
\begin{align}
\hat{\mathcal{H}}_\mathrm{mech} = \frac{1}{2}\Bigl(\tfrac{\hat{p}^2}{m} + m\omega_\mathrm{m}^2\, \hat{x}^2\Bigr),
\end{align}
where $m$ is the mirror mass and $\omega_\mathrm{m}$ is the mechanical resonant frequency. Consequently, in the Markovian limit where $\kappa_c$ is sufficiently larger than $\kappa$, $\Gamma$, and $J$, the system’s Langevin equations of motion in the time domain are given by:
\begin{align}
    \frac{\mathrm{d}}{\mathrm{d}t}\hat{A} &= \sqrt{2\kappa}\,\hat{a}_{\mathrm{in}} - (\kappa+\Gamma)\,\hat{A} + (iJ+\Gamma)\,\hat{B} \nonumber\\
    &\quad + iJ^\prime\sqrt{\frac{2}{\kappa_c}}\,\hat{c}_{\mathrm{in}} - i\alpha\,\hat{x}, \nonumber\\
    \frac{\mathrm{d}}{\mathrm{d}t}\hat{B} &= \sqrt{2\kappa}\,\hat{b}_{\mathrm{in}} - (iJ-\Gamma)\,\hat{A} - (\kappa+\Gamma)\,\hat{B} \nonumber\\
    &\quad - iJ^\prime\sqrt{\frac{2}{\kappa_c}}\,\hat{c}_{\mathrm{in}} + i\alpha\,\hat{x}, \nonumber\\
    \hat{C} &= \frac{2}{\kappa_c}\,\hat{c}_{\mathrm{in}} - i\frac{J^\prime}{\kappa_c}\bigl(\hat{A}^\dagger - \hat{B}^\dagger\bigr), \nonumber\\
    \frac{\mathrm{d}}{\mathrm{d}t}\hat{x} &= \frac{\hat{p}}{m}, \nonumber\\
    \frac{\mathrm{d}}{\mathrm{d}t}\hat{p} &= -m\omega_\mathrm{m}^2\,\hat{x} \;-\; \hbar\,\alpha\,\bigl[(\hat{A}+\hat{A}^\dagger) - (\hat{B}+\hat{B}^\dagger)\bigr] + F, \label{eqs:EoM}
\end{align}
where $\hat{a}_\mr{in},\ \hat{b}_\mr{in}$ and $\hat{c}_{\mathrm{in}}$ are the input fields to $\hat{A},\ \hat{B}$ and $\hat{C}$, respectively. Here, $\kappa$ denotes the damping rate of the mode $\hat{A}$ and $\hat{B}$, $\kappa_c$ denotes that of the mode $\hat{C}$, and we define $\Gamma \equiv |J^\prime|^2/\kappa_c$ as the effective tunneling rate between cavities $\hat{A}$ and $\hat{B}$ via $\hat C$. Because $\kappa_c$ is sufficiently large, the dynamics of the mode $\hat{C}$ can be adiabatically eliminated. By choosing
\begin{align}
    J = i\Gamma,
\end{align}
we obtain unidirectional coupling from $\hat{A}$ to $\hat{B}$, enabling the speed meter--type coupling described in the previous section.

Hereafter, we use the two-photon formalism (see, e.g., Refs.~\cite{PhysRevA.31.3093,PhysRevA.31.3068}). In this formalism, the input and output fields are represented in quadrature amplitudes as 
\[
\pmb{k}_{\Omega} = \{k_{1,\Omega}, k_{2,\Omega}\}^\mathrm{T},
\]
for $\hat{k}\in(\hat{a},\hat{b},\hat{c})$ where, after Fourier transformation, the components are defined as
\begin{align*}
    k_{1,\Omega} &= \frac{k_{\omega_0+\Omega}+k^\dagger_{\omega_0-\Omega}}{\sqrt{2}}, \\
    k_{2,\Omega} &= \frac{k_{\omega_0+\Omega}-k^\dagger_{\omega_0-\Omega}}{i\sqrt{2}},
\end{align*}
with $\omega_0$ as the carrier frequency and $\Omega$ as the sideband frequency. For brevity, we omit the explicit $\Omega$ dependence and instead use the subscripts ``in'' and ``out'' to denote input and output fields (e.g., $\pmb{k}_{\mathrm{out}} = \{k_{1,\mathrm{out}}, k_{2,\mathrm{out}}\}^\mathrm{T}$).
For the cavity field \(X=(\hat{A},\hat{B},\hat{C})\), we define the quadratures after Fourier transform as
\begin{align*}
    X_{1,\Omega} = \frac{X_\Omega+X_\Omega^\dagger}{\sqrt{2}},\quad
    X_{2,\Omega} = \frac{X_\Omega-X_\Omega^\dagger}{i\sqrt{2}},
\end{align*}
and denote \(\pmb{X}=\{X_{1,\Omega}, X_{2,\Omega}\}^T\).
The output field from mode $\hat B$ is given by
\[
\pmb{\hat{b}}_{\mathrm{out}} = -\pmb{\hat{b}}_{\mathrm{in}} + \sqrt{2\kappa}\,\pmb{\hat{B}}.
\]
By solving Eqs.~\eqref{eqs:EoM} in the frequency domain, under the assumptions that $\omega_\mathrm{m}\rightarrow0$ (a free mass), the output field from mode $\hat B$ is obtained as
\begin{align}
    \pmb{\hat{b}}_{\mathrm{out}} = 
    \tq[d]{D}\,\pmb{\hat{a}}_{\mathrm{in}} +
    \tq[e]{D}\,\pmb{\hat{b}}_{\mathrm{in}} +
    \tq[c]{D}\,\pmb{\hat{c}}_{\mathrm{in}} +
    \pmb{t}_F\,\frac{F}{F_\mathrm{SQL}}, \label{eq:d2out}
\end{align}
with the coefficient matrices defined as
\begin{align}
     \tq[\mathit{d}]{D} &=  \matr{-\frac{1}{(i+\omega)^2}}{0}{-\frac{\gamma}{(i+\omega)^4}}{-\frac{\omega}{(i+\omega)^2}}, \label{eq:D1}\\
    \tq[\mathit{e}]{D} &= \matr{-\frac{\omega}{i+\omega}}{0}{\frac{\gamma}{\omega(i+\omega)^3}}{-\frac{\omega}{i+\omega}}, \label{eq:D2}\\
    \tq[\mathit{c}]{D} &= \matr{0}{\frac{i\omega}{(i+\omega)^2}}{-\frac{i\omega}{(i+\omega)^2}}{\frac{\gamma(i+2\omega)}{\omega(i+\omega)^4}}, \label{eq:C}\\
\pmb{t}_F &= \vect{0}{-\frac{i\sqrt{2\gamma}}{(i+\omega)^2}}, \label{eq:F}
\end{align}
and given the following parametrization
\begin{equation}
\alpha \equiv \sqrt{\frac{\Theta m}{2\hbar}}, \quad
\gamma \equiv \frac{\Theta}{8\kappa^3}, \quad
\omega \equiv \frac{\Omega}{2\kappa}.
\end{equation}
where $\Theta$ is the normalized pump power.
Here, the force SQL is defined as:
\begin{align}
    F_\mr{SQL} = \sqrt{2\hbar m\Omega^2}.
\end{align}
Detection is performed by projecting the output field onto the homodyne vector
\begin{align}
\pmb{H}_{\phi}^T = \{\sin\phi, \cos\phi\}:
\end{align}
as
\begin{align}
\hat{b}_\phi = \pmb{H}_{\phi}^T\,\pmb{\hat{b}}_{\mathrm{out}}.
\end{align}
The force spectral density is then
\begin{align}
    S^F(\Omega) = F_\mathrm{SQL}^2\,\frac{\sum_{\mu} \pmb{H}_{\phi}^T\,\tq[\mu]{D}\,\tq[\mu]{S}^\mathrm{in}\,\tq[\mu]{D}^\dagger\,\pmb{H}_{\phi}}{\bigl|\pmb{H}_{\phi}^T\,\pmb{t}_F\bigr|^2}, \label{eq:SF}
\end{align}
where $\mu\in(d,e,c)$. The quantum spectral density of the input fields, $\tq[\mu]{S}^\mathrm{in}$, is defined by
\begin{align}
    2\pi\,&\delta(\Omega-\Omega')\,\tq[\mu,\mathit{ij}]{S}^\mathrm{in}(\Omega) \notag\\
    &\equiv \tfrac{1}{2}\langle \mathrm{in}| \hat{b}_{i,\mr{in}}(\Omega)\,\hat{b}_{j,\mr{in}}^\dagger(\Omega') + \hat{b}_{j,\mr{in}}^\dagger(\Omega')\,\hat{b}_{i,\mr{in}}(\Omega)|\mathrm{in}\rangle,
\end{align}
where $(i,j)=\{1,2\}$ and $|\mathrm{in}\rangle$ is the quantum state of the input fields (see Refs.~\cite{Danilishin_2012,2018LSA.....7...11D} for more details).

\section{Force sensitivity}\label{sec:IV}
The system’s response function to an external force is given by $\pmb{t}_F$ in Eq.~\eqref{eq:F} and exhibits a speed meter--type behavior where the frequency response scales as $1/\omega$ at low frequencies ($\Omega\ll\kappa$). In this regime, the off-diagonal term in $\tq[\mathit{d}]{D}$ in Eq.~\eqref{eq:D1} (associated with radiation pressure from $\pmb{\hat{a}}_{\mathrm{in}}$) remains constant. This property allows one to cancel out the term by correlating the phase and amplitude quadrature noises via balanced homodyne detection at a fixed angle. The optimal homodyne angle at DC is
\begin{align}
    \phi_\mathrm{opt} = \tan^{-1} \gamma,
\end{align}
where the off-diagonal term in Eq.~\eqref{eq:D1} is fully canceled.

Examining the noise contributions in Eqs.~\eqref{eq:D2} and \eqref{eq:C} from $\pmb{\hat{b}}_{\mathrm{in}}$ and $\pmb{\hat{c}}_{\mathrm{in}}$, respectively, reveals two important points. First, when the mechanics is turned off ($m\rightarrow\infty$), the remaining terms vanish at low frequencies ($\Omega\ll\kappa$). These residual terms, also discussed in Eq.~(32) of Ref.~\cite{PhysRevX.5.021025}, can be viewed as a drawback of creating nonreciprocal interactions using only reciprocal interactions.

Second, the radiation pressure terms—namely, the off-diagonal entry in Eq.~\eqref{eq:D2} and the diagonal entry in Eq.~\eqref{eq:C}, are both proportional to $\gamma$ and do not exhibit speed meter--type behavior; in fact, they scale as $\omega^{-1}$ at $\omega\ll1$, implying that they dominate the speed meter--type noise in Eq.~\eqref{eq:D1}. This issue arises from loop noise generated by coupling $\hat{x}$ to the cavity modes $\hat{A}$ and $\hat{B}$, through which the three output fields get entangled. Interestingly, this unwanted contribution can be canceled by detecting the output fields $\hat{a}_{\mathrm{out}}$ and $\hat{c}_{\mathrm{out}}$ at fixed, appropriate angles and then feeding them forward to $\hat{b}_{\mathrm{out}}$ via Wiener filters (see Appendix~\ref{sec.a.2}). Because these detection angles are fixed, no additional filter cavities are required, thereby avoiding extra infrastructure.

After filtering, the output $\hat{b}_{\mathrm{opt}}^\mr{(c)}$ is given by
\begin{widetext}
\begin{align}
    \hat{b}_{\mathrm{opt}}^\mr{(c)} = \frac{1}{\sqrt{\gamma^2+1}}
    \Biggl\{
    -\frac{\gamma\,\omega^2\,(\omega^2+2)}{(\omega + i)^2}\,\hat{a}_{1,\mathrm{in}}
    -\frac{1}{(\omega + i)^2}\,\hat{a}_{2,\mathrm{in}}
    -\frac{\omega}{\omega + i}\,\hat{b}_{2,\mathrm{in}}
    -\frac{i\omega}{(\omega + i)^2}\,\hat{c}_{1,\mathrm{in}}
    -\frac{i\sqrt{2\gamma}}{(\omega + i)^2}\,\frac{F}{F_\mathrm{SQL}}
    \Biggr\},
    \label{eq:x2out}
\end{align}
\end{widetext}
Here $\gamma$ represents the internal laser power, and both $\gamma$ and $\omega$ are dimensionless. From Eq.~\eqref{eq:x2out}, the force spectral density becomes
\begin{align}
\frac{S^F_\mathrm{SM}}{S^F_\mathrm{SQL}}
= \frac{1}{2}\Bigl[\frac{(1+\omega^2)^2}{\gamma} + \gamma\,\omega^4(2+\omega^2)\Bigr],\label{eq:speed}
\end{align}
where $S^F_\mathrm{SM}$ and $S^F_\mathrm{SQL}(=F_\mr{SQL}^2)$ denote the spectral densities of the speed meter and the standard quantum limit, respectively.

\begin{figure}
    \centering
    \includegraphics[width=1.0\linewidth]{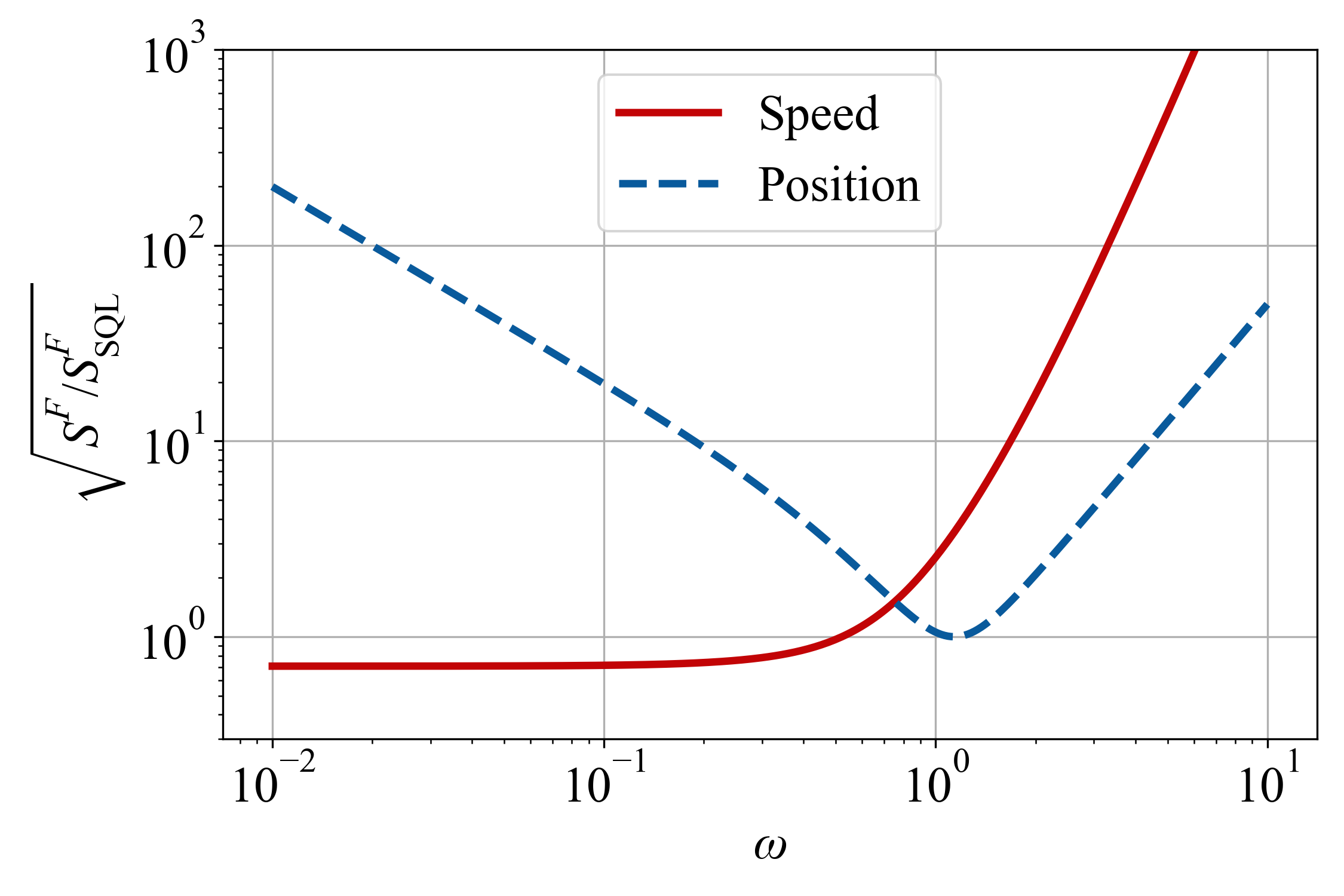}
    \caption{Force spectral densities for the speed meter (red solid) and the position meter (blue dashed) with $\gamma=1$, each normalized by $S_\mathrm{SQL}^F$, the standard quantum limit of force measurement. $\omega$ is the normalized frequency defined as $\omega\equiv\Omega/2\kappa$.}
    \label{fig:SensitivitySpeedmeter}
\end{figure}

For comparison, we also compute the noise spectral density of a position meter that employs only cavity $\hat{A}$. To ensure a fair comparison, the coupling strength $\alpha$ is doubled relative to that of the speed meter, and the bandwidth $\kappa$ remains the same. In the low-frequency regime, the spectrum is
\begin{align}
\frac{S^F_\mathrm{PM}}{S^F_\mathrm{SQL}}
= \frac{1}{2}\Bigl[\frac{\omega^2\,(2+\omega^2)}{8\gamma} + \frac{8\gamma}{\omega^2\,(2+\omega^2)}\Bigr], \label{eq:position}   
\end{align}
(see Appendix~\ref{sec.a.1} for detailed derivations). The complete spectral dependencies for both the speed meter and the position meter configurations are shown in Fig.~\ref{fig:SensitivitySpeedmeter}. The sensitivity of the speed meter shown in red curve is better than that of the position meter almost at low frequencies $\omega<1$, due to the improvement of the radiation pressure noise. 

\section{Discussion}\label{sec:discussion}

The velocity measurement feature manifests at low frequencies, as evidenced by Eq.~\eqref{eq:speed}. In the limit $\omega\rightarrow0$, the noise spectral density converges to a constant value:
\[
\frac{S^F_\mathrm{SM}}{S^F_\mathrm{SQL}} = \frac{1}{2\gamma},
\]
which implies that increasing the pump laser power (i.e., increasing $\gamma$) can reduce the noise arbitrarily. This noise reduction can also be achieved by injecting a squeezed state of light into cavity $\hat{A}$ rather than a thermal vacuum, which is consistent with the behavior observed in previously proposed speed meter configurations~\cite{PhysRevD.66.122004, Chen_2003, 2018LSA.....7...11D}.

In contrast, the low-frequency noise for the position meter, as given by Eq.~\eqref{eq:position}, scales as
\[
\frac{S^F_\mathrm{PM}}{S^F_\mathrm{SQL}} = \frac{4\gamma}{\omega^2},
\]
indicating that this noise increases with the laser power. Our analysis does not take into account imperfections such as optical losses in the cavities or non-Markovian correlations that arise when $\kappa_c$ is not arbitrarily large. These effects are left for future work.

A notable feature of our approach is that it realizes velocity measurement using only reciprocal interactions. For example, in the gravitational-wave detection community, Sagnac-type speed meters based on ring cavities and polarization-based speed meters have been proposed~\cite{Chen_2003,2018LSA.....7...11D}. These configurations utilize either the spatial or polarization degrees of freedom for velocity measurement. However, utilizing spatial degrees of freedom is constrained by the detector infrastructure, while polarization-based schemes encounter challenges such as birefringence and intensity-to-phase coupling—issues that require further research.

\begin{figure}
    \centering
    \includegraphics[width=0.9\linewidth]{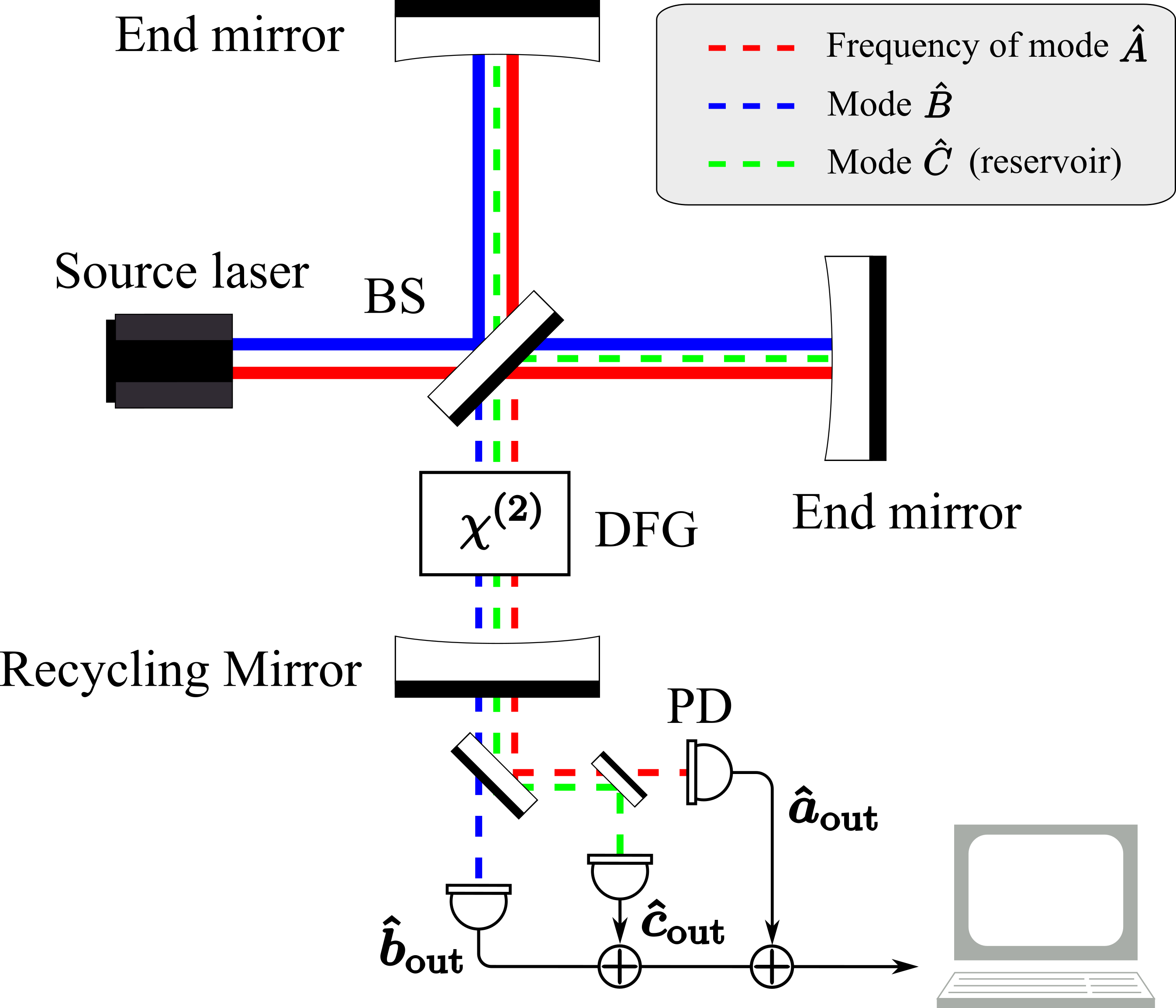}
    \caption{Reservoir engineering in a Michelson interferometer. Bright fields from the source laser enter from the east side and pump the interferometer arms, mediating coupling between the mechanical mode $\hat{x}$ and the cavity modes $\hat{A}$ and $\hat{B}$. The south port of the beam splitter is kept dark, allowing only sidebands generated by differential motion of the end mirrors to exit. The end mirrors and the recycling mirror together form the differential cavity mode of the interferometer. Abbreviations: BS, beam splitter; PD, photodetector (homodyne detector); DFG, differential frequency generation.}
    \label{fig:Michelson}
\end{figure}

In contrast, our method relies solely on reciprocal interactions and, in its simplest implementation, can potentially be realized by combining three cavities. In this sense, it is similar to a sloshing-type speed meter~\cite{PhysRevD.66.122004}. One possible approach is to insert a nonlinear optical crystal into a single cavity and pump it with light at three appropriate frequencies to generate the three beam-splitter interactions. Fig.~\ref{fig:Michelson} illustrates an implementation in a Michelson interferometer. The three cavity modes, each at a different frequency, are spatially degenerate within a single cavity formed by the end mirror and the recycling mirror (representing the differential mode of the interferometer). The nonlinear crystal inserted into this cavity mediates beam-splitter interactions among the three modes---labeled $\hat{A}$, $\hat{B}$, and $\hat{C}$---via differential frequency generation (DFG). The output fields are separated using dichroic mirrors, measured via homodyne detection, and then combined using optimal filters. To satisfy the Markovian limit, the signal extraction rate of the reservoir mode $\hat{C}$ must be sufficiently high; this requirement can be met by appropriately designing the coating parameters of the recycling mirror.


\section{Conclusion}\label{sec:conclusion}

In this work, we have presented a new approach to back-action evasion using reciprocal interactions. By employing an engineered reservoir in a multi-cavity system combined with conditional feed-forward, we achieve an effective nonreciprocal coupling and QND speed measurement. Our analysis shows that the force noise spectral density of the resulting speed meter converges to a constant value at low frequencies, in contrast to conventional position meters, whose noise increases as the frequency decreases.

An advantage of this approach is that the measurement noise can be reduced by increasing the pump laser power or by injecting a squeezed state into cavity $\hat{A}$, allowing performance to approach or even exceed the standard quantum limit. Moreover, compared to other speed meter proposals, our method relies solely on reciprocal interactions, which avoids some of the technical challenges involved in engineering nonreciprocal interactions in the spatial or polarization domains.

For future work, experimental validation and further optimization will be necessary to assess the practical advantages of the proposed scheme. In the context of gravitational-wave detectors, this includes detailed studies of optical loss, mode matching, and mirror coating properties. Our results suggest that velocity measurement based solely on reciprocal interactions provides an alternative approach for high-precision sensing.

\begin{acknowledgments}
Research by Y. N. is supported by JSPS Grant-in-Aid for JSPS Fellows Grant Number 23KJ0787 and 23K25901. J.W.G. is supported by the Australian Research Council Centre of Excellence for Gravitational Wave Discovery (Project No. CE170100004 and CE230100016), an Australian Government Research Training Program Scholarship, and also partially by the US NSF grant PHY-2011968. In addition, Y.C. acknowledges the support by the Simons Foundation (Award Number 568762).  This work was partially supported by JST ASPIRE (JPMJAP2320). 
\end{acknowledgments}

\appendix

\section{Full input-output relation and Wiener filtering}\label{sec.a.2}
\begin{figure}
    \centering
    \includegraphics[width=0.9\linewidth]{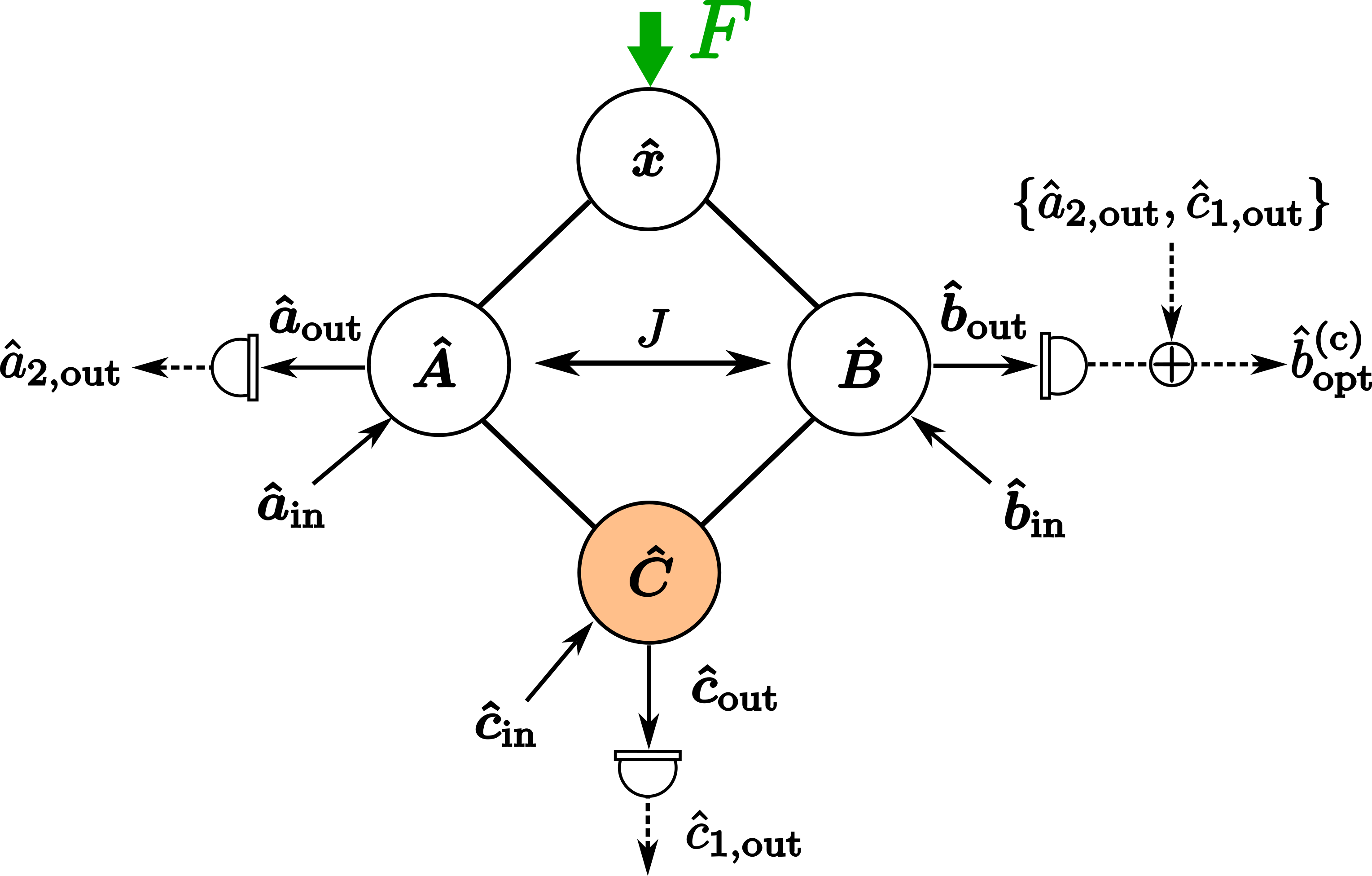}
    \caption{Mode diagram for the full input-output relation of our scheme.}
    \label{fig:IO}
\end{figure}
We now derive the full multi-input, multi-output (MIMO) relation for our scheme and discuss the implementation of conditional feed-forward, following an approach similar to that in Ref.~\cite{PhysRevA.110.022601}.
From Eqs.~\ref{eqs:EoM} and the standard input-output relations:
\begin{align}
    \pmb{\hat{a}}_{\mathrm{out}} &= -\pmb{\hat{a}}_{\mathrm{in}} + \sqrt{2\kappa}\,\pmb{\hat{A}}, \notag \\
    \pmb{\hat{c}}_{\mathrm{out}} &= -\pmb{\hat{c}}_{\mathrm{in}} + \sqrt{2\kappa_c}\,\pmb{\hat{C}}. \notag
\end{align}
we obtain the input-output relation (with Eq.~\ref{eq:d2out}) as follows:
\begin{widetext}

\begin{align}
    \pmb{\hat{a}}_\mr{out} &= \matr{-\frac{\omega}{i+\omega}}{0}{\frac{\gamma}{\omega(i+\omega)^3}}{-\frac{\omega}{i+\omega}}\pmb{\hat{b}}_\mr{in} 
    + \matr{0}{0}{-\frac{\gamma}{\omega^2(i+\omega)^2}}{0} \pmb{\hat{a}}_\mr{in}
    + \matr{0}{-\frac{i}{i+\omega}}{\frac{i}{i+\omega}}{-\frac{\gamma(i+2\omega)}{\omega^2(i+\omega)^3}}\pmb{\hat{c}}_\mr{in}
    + \vect{0}{\frac{i\sqrt{2\gamma}}{\omega(i+\omega)}}\frac{F}{F_\mr{SQL}} \\
    \pmb{\hat{c}}_\mr{out} &= \matr{-\frac{\gamma(i+2\omega)}{\omega(i+\omega)^4}}{-\frac{i\omega}{(i+\omega)^2}}{\frac{i\omega}{(i+\omega)^2}}{0}\pmb{\hat{b}}_\mr{in} 
    + \matr{\frac{\gamma(i+2\omega)}{\omega^2(i+\omega)^3}}{\frac{i}{i+\omega}}{-\frac{i}{i+\omega}}{0} \pmb{\hat{a}}_\mr{in}
    + \matr{\frac{\omega^2}{(i+\omega)^2}}{\frac{\gamma(i+2\omega)^2}{\omega^2(i+\omega)^4}}{0}{\frac{\omega^2}{(i+\omega)^2}}\pmb{\hat{c}}_\mr{in}
    + \vect{0}{-\frac{i\sqrt{2\gamma}(i+2\omega)}{\omega(i+\omega)^2}}\frac{F}{F_\mr{SQL}}.
\end{align}
    
\end{widetext}

To subtract the excess back-action noise, we need information on the quadrature $\hat{x}_\mr{1,in}$ and $\hat{p}_\mr{c,in}$. This can be achieved by choosing the homodyne angles as:
\begin{align*}
    \hat{a}_{2,\mr{out}} &= \pmb{H}_{0}^T\,\pmb{\hat{a}}_{\mathrm{out}}, \\
    \hat{c}_{1,\mr{out}} &= \pmb{H}_{\pi/2}^T\,\pmb{\hat{c}}_{\mathrm{out}}.
\end{align*}
Subtracting these with optimal filters $g_{\{1,2\}}$ defined as:
\begin{align}
    g_1 &= \frac{i\gamma [1-\omega^2(i+\omega)^2]}{\sqrt{1+\gamma^2}\omega(i+\omega)^2}, \\
    g_2 &= \frac{\gamma(1-\omega^2-\omega^4)}{\sqrt{1+\gamma^2}\omega(i+\omega)},
\end{align}
the conditional-filtered state is obtained as:
\begin{align}
    \hat{b}_{\mathrm{opt}}^\mr{(c)} = \hat{b}_{\mathrm{opt}} - g_1\, \hat{a}_{2,\mr{out}}- g_2\, \hat{c}_{1,\mr{out}}.
\end{align}
whose breakdown is shown in Eq.~\eqref{eq:x2out}. This cancels the loop noise as required.

\section{Position Meter}\label{sec.a.1}
We analyze the sensitivity of the position as shown in Fig.~\ref{fig:SensitivitySpeedmeter}. In this configuration, the optical mode $\hat A$ is coupled to the mechanical mode \(\pmb{\hat{x}}\), while \(\pmb{\hat{a}}_\mr{in}\) and \(\pmb{\hat{a}}_\mr{out}\) represent the incoming and outgoing optical fields, respectively. The system Hamiltonian is given by
\begin{align}
    \hat{H} = \hbar \alpha (\hat{A}+\hat{A}^\dagger)\left(\hat{x}-\frac{h\,L}{2}\right)
    + \frac{1}{2}\left(\frac{\hat{p}^2}{m}+m\,\omega_\mr{m}^2\,\hat{x}^2\right).
\end{align}
In the quadrature representation, the equations of motion become:
\begin{align}
    -i\Omega \hat{A}_1 &= -\gamma\, \hat{A}_1 + \sqrt{2\gamma}\,\hat{a}_{1,\mr{in}}, \notag\\[1mm]
    -i\Omega \hat{A}_2 &= -\gamma\, \hat{A}_2 + \sqrt{2\gamma}\,\hat{a}_{2,\mr{in}}
    -\sqrt{2}\,\alpha\Bigl(\hat{x}-\frac{hL}{2}\Bigr), \notag\\[1mm]
    -i\Omega \hat{x} &= \frac{\hat{p}}{m}, \notag\\[1mm]
    -i\Omega \hat{p} &= -\sqrt{2}\,\hbar\,\alpha\,\hat{A}_1 - m\,\omega_\mr{m}^2\,\hat{x}\,. 
\end{align}

Using the standard input--output relation,
\[
\pmb{\hat{a}}_\mr{out} = -\pmb{\hat{a}}_\mr{in}+\sqrt{2\gamma}\,\pmb{\hat{A}},
\]
the transfer matrix and signal vector for the Michelson interferometer are found to be
\begin{align}
    \tq[pm]{T} = \begin{pmatrix}
        -1+\frac{2i}{i+2\omega} & 0 \\
        \frac{8\gamma}{\omega^2(i+2\omega)^2} & -1+\frac{2i}{i+2\omega}
    \end{pmatrix},
\end{align}
\begin{align}
    \pmb{t}_{F,\mr{pm}} = \begin{pmatrix}
        0 \\
        \frac{4i\sqrt{\gamma}}{\omega(i+2\omega)}
    \end{pmatrix},
\end{align}
where $\mathcal{K}_\mr{pm} = \frac{4\gamma\hbar\,\alpha^2}{m\Omega^2(\kappa^2+\Omega^2)}$.
Thus, the output field can be expressed as
\begin{align}
    \pmb{\hat{a}}_\mr{out} = \tq[pm]{T} \pmb{\hat{a}}_\mr{in} + \frac{F}{F_\mr{SQL}} \pmb{t}_{F,\mr{pm}}.
\end{align}
Calculating the force spectral density using Eq.~\eqref{eq:SF}, we then obtain Eq.~\eqref{eq:position}.